\documentclass[a4paper,10pt,twocolumn]{article}
\usepackage{graphicx}

\newcommand{\be}{\begin{equation}}
\newcommand{\ee}{\end{equation}\noindent}
\newcommand{\bea}{\begin{eqnarray}}
\newcommand{\eea}{\end{eqnarray}}

\newcommand{\beqa}{\begin{eqnarray}} 
\newcommand{\eeqa}{\end{eqnarray}}
\usepackage{graphicx, color}
\usepackage{latexsym}
\newcommand{\vect}[1]{\mbox{\boldmath${#1}$}}
\newcommand{\vex}{{\vect x}}

\begin{document}

\title{
Transport coefficients of \\
Relativistic Causal Hydrodynamics for Hadrons
}
\author{Shin Muroya \\
\qquad 
{\small (Dept.\ of C.\ M., Matsumoto University, }\\
{\small Niimura, Matsumoto, Nagano 390-1295 Japan ) 
} }

\date{}
\vspace{.5cm}
\setlength{\textwidth}{16.cm}

\twocolumn[
    \maketitle

\begin{small}
{\bf Abstract}
We investigate coefficients in the Israel-Stewart's causal 
hydrodynamics and discuss the way to calculate them microscopic 
theory. Based on the hadro-molecular simulation based on an 
event generator URASiMA, we evaluate the coefficients for a hot 
and dense hadronic fluid.\vspace{1.cm}


\end{small}
\vspace{1.cm}

]

\setlength{\textwidth}{16.5cm}

\noindent \section{
{\bf Introduction}
}

A hydrodynamical model is one of the most established models for the 
multiple-production in high energy reactions. Especially for RHIC, a 
hydrodynamical model is believed to be the promising model for 
the $v_2$.   Up to now, most hydrodynamical models for RHIC adopt a 
perfect fluid model, i.e., Euler equation neglecting viscosity and 
heat conductivity for simplicity.

It is well know that the relativistic extension of the Navier-Stokes 
equation which contains transport coefficients is not unique.  At least 
three types of deferent definition for the local four velocity 
$U^{\mu}$ are known. In the textbook by Landau and Lifshitz, 
$U^{\mu}$ is defined as an eigenvector of the energy-momentum tensor 
$^{\cite{LL}}$, and it is named e-frame.  A mass current $J^{\mu}$ is 
used to define $U^{\mu}$ by Eckart and 
Namiki-Iso$^{\cite{NI} }$(this is named m-frame). Weinberg used a charged 
current instead of a mass current$^{\cite{We}}$. All these three types 
are the covariant extension of the non-relativistic Navier-Stokes 
equation which belongs to the diffusion type parabolic equation. 
The rank of the spatial derivative and the one of time derivative 
are different. This difference between space and time contradicts 
relativity and can lead acausal results.

Israel and Stewart proposed a relativistic causal hydrodynamic 
equation$^{\cite{IS}}$ which is hyperbolic type and contains several 
additional coefficients other than ordinary transport coefficients. 
In this paper we will investigate the physical meaning of these 
new coefficients and evaluate them of a hadronic gas based on a 
microscopic theory.

\noindent 
\section{
{\bf Causal Hydrodynamics }
}
The relativistic hydrodynamical equation proposed by Israel and Stewart is as follows;
\be
\begin{array}{lcl}
q^{\mu} &=& -\kappa \Delta^{\mu \nu}
(\displaystyle{\frac{1}{T}} \partial _{\nu} T + D u_{\nu}  \\
&& \left.  + \bar{\beta_{1}} Dq_{\nu}
-  \bar{  \alpha _{0}} \displaystyle{ \partial_{\nu}\sigma} -
   \bar{  \alpha _{1}} \partial_{\alpha}\sigma^{\alpha}_{\nu} \right)
\nonumber \\
\noalign{\vskip 0.4cm}
\sigma &=& -\displaystyle{\frac{1}{3}}\eta_{V}(\partial _{\mu} u^{\mu} 
+  \beta_{0} D\sigma -  \bar{\alpha_{0}}\partial_{\mu}q^{\mu})  
\nonumber \\
\noalign{\vskip 0.4cm}
\sigma_{\mu \nu} &=& -2 \eta < \partial_{\mu}u_{\nu}
+  {\beta_{2}} D\sigma_{\mu \nu}
-  \bar{  \alpha _{1}}\partial_{\mu}q_{\nu}>
\nonumber 
\end{array}
\nonumber 
\ee
where 
$
D  \equiv  u_{\mu}\partial^{\mu}$ and $
<A_{\mu \nu}>  \equiv  \displaystyle{\frac{1}{2}}
\Delta_{\mu}^{\lambda}\Delta_{\nu}^{\rho}(A_{\lambda\rho}+A_{\rho\lambda}
-\displaystyle{\frac{2}{3}}
\Delta_{\lambda \rho}\Delta^{\alpha \beta}A_{\alpha \beta}) 
$

The most important feature of Israel-Stewart's hydrodynamics is the 
existence of the relaxation terms, $\beta_{i}$, which makes the equation 
hyperbola.  As the same order term, there also appear terms with 
$\alpha_i$ which stand for the influence of the derivative of the 
different currents.

\section{Linear Responce Theory}

A hydrodynamical model is a macroscopic phenomenological model of which coefficients should be derived by a microscopic statistical physics.  According to the linear response theory, transport coefficients are evaluated from the correlations of the corresponding currents. Thermo-dynamical forces and currents are identified through the formula of entropy production.

Israel and Stewart gives entropy production as$^{\cite{IS}}$,
\be
\begin{array}{lcl}
\partial_{\mu}S^{\mu} 
&=& \displaystyle{\frac{1}{T} }q_{\mu} \quad
\{-\Delta^{\mu \nu}
(\displaystyle{\frac{1}{T}} \partial _{\nu} T + D u_{\nu} + \bar{\beta_{1}}
 D{ q_{\nu}} \\ \noalign{\vskip 0.2cm}
&& \qquad \qquad -{ \bar{ \alpha _{0}}} \partial_{\nu}{ \sigma} -
 { \bar{ \alpha _{1}}} \partial_{\alpha}{ \sigma^{\alpha}_{\nu}}  ) \}
\\ \noalign{\vskip 0.2cm}
&+&
\displaystyle{\frac{1}{T}}\sigma \quad \left(
-(\partial _{\mu} u^{\mu} 
 +{ \beta_{0}} D{ \sigma} - { \bar{\alpha_{0}}}\partial_{\mu}
{ q^{\mu}} )
\right)  \nonumber \\
\noalign{\vskip 0.4cm}&+&
\displaystyle{\frac{1}{T}}\sigma_{\mu\nu}\left( - < \partial_{\mu}u_{\nu} 
+ { \bar{\beta_{2}}} D{ \sigma_{\mu \nu}}
- { \bar{ \alpha _{1}}}\partial_{\mu}{ q_{\nu}}> \right)  
\\ \noalign{\vskip 0.2cm}
&=& 
\displaystyle{\frac{q_{\mu}q^{\mu}}{\kappa T} }
+ 
\displaystyle{\frac{3}{\eta_{V}} \displaystyle{\frac{\sigma ^2}{T}}
+ 
\displaystyle{\frac{ \sigma_{\mu\nu} \sigma^{\mu\nu} } {2\eta T} }  }. 
\end{array} \nonumber
\ee
This formula is diagonal with currents and there exists no combination 
which contradicts with Currie's theorem.  Therefore, even there exist 
$\alpha$ term, terms added by Israel and Stewart are able to be rewrite 
in the higher derivative of thermo-dynamical quantities.  Under the basic 
assumption of linear response theory that the disturbance of macroscopic 
current is small, we can safely adopt usual Kubo formula for the 
transport coefficients, such as viscosity and heat conductivity\cite{MS}.

\section{Evaluation of $\beta$ }

As we have discussed in the previous section, we can calculate viscosity 
and heat conductivity by taking current-current correlation as usual.  
If we use Israel-Stewart's causal hydrodynamics as a phenomenological 
model, we need additional coefficients $\alpha$ and $\beta$ also.  
Neglecting $\alpha$ for simplicity, let us focus our discussion on 
$\beta$ through out this paper.

If the $\beta$ belongs to a kind of material constant, it should only depend on the thermo-dynamical quantities such as temperature and density.  It should not depend on the boundary condition of the macroscopic current.  Hence, we may assume particular situation without generality. 

Suppose the situation that the temperature gradient only exists.  Then thermal current $q^{i}$ in eq.\ (1) is given as,
\be
q^{i} = -\kappa 
 { \bar{\beta_{1}}} \displaystyle{\frac{d}{dt}}q^{i}.  \nonumber \\
\ee
If the heat current relaxes in $\tau_{\kappa}$ as, 
\be
q^{i}(t) = q^{i}(0)e^{-\frac{t}{\tau_{\kappa}}}. 
\ee
Then, $\bar{\beta_{1}}$ are related to the heat conductivity as,
$
\kappa  \bar{\beta_{1}} = \tau_{\kappa}. 
$

According to the linear response theory, heat conductivity is calculated through the correlation of the heat currents,
\be
\kappa = \frac{1}{T}\int d^3 \vex' \int^{t}_{-\infty} dt' {\rm e}^{-\varepsilon(t-t')}(q_{x}(\vex,t),q_{x}(\vex',t')). 
\ee
Therefore, if the heat current relaxes in the relaxation time $\tau_{\kappa}$, heat conductivity is written as,
$$
\kappa = \frac{\tau_{\kappa}}{T} <<q^{i}(t)q^{i}(t)>>,
$$
where $<< ** >>$ is average and integraion in eq.(5).

Hence, transport coefficient and corresponding $\beta$ are closely related through the expectation value of the square of the current, relaxation time and temperature.

\section{Relaxation of hadron gas }

\begin{figure}[bh]
\includegraphics[width=.98 \linewidth]{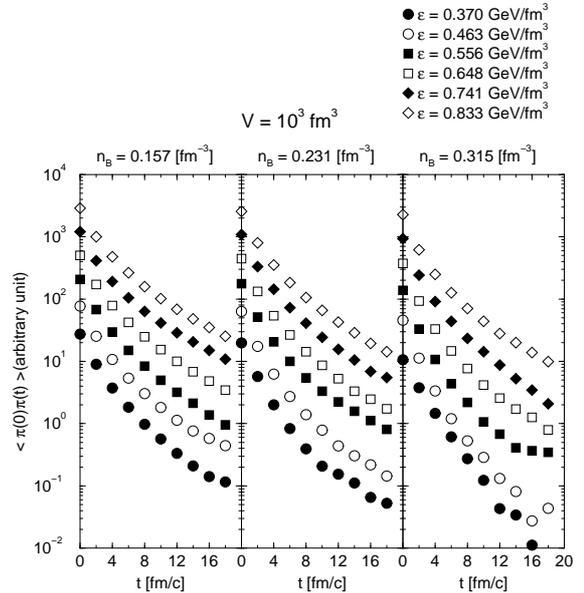}

\caption{Relaxation in visous-shear tensor correlation}
\end{figure}

In a previous paper we have reported the calculation of transport coefficients of hadron gas based on a hadro-molecular simulation$^{\cite{MS}}$.  We used event generator URASiMA as a time evolution generator.  In our simulation, we first put only baryons in the box with periodic boundary condition and turn switch on the simulator.  Collisions between particles take place and mesons and resonances are produced.  At later time, the system approaches to some kind of stationary state where all kinds of particle possess a common slope parameter, which we may call "temperature"$^{\cite{Sasaki-ptp}}$. 
The configurations of the stationary state we use as statistical ensembles.

Figure 1 displays correlation of stress-shear tensors as a function of time interval of the currents, which correspond to $(T_{xy}(\vex,t),T_{xy}(\vex',t'))$ in the formula of the linear response theory.  Figures 2 and 3 display the shear viscosity and heat conductivity of hadron gas obtained by our simulation. According to our simulation, both transport coefficients depend on baryon number density very weakly and temperature dependences are as strong as almost $T^5$$^{\cite{MS}}$.

On the other hand, the changes of the relaxation times of the currents are very mild as functions of temperature (figs.\ 4 and 5).  Most part of the temperature dependences of the transport coefficients come from the change of the intensity (expectation value of the current square (figs.\ 6 and 7)).

\begin{figure}[hbt]
\hspace{-0.5cm}
\includegraphics[width=1. \linewidth]{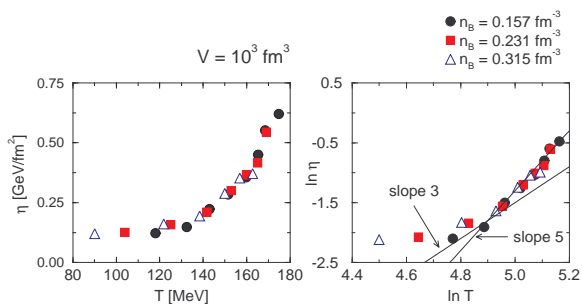}
\caption{Shear visosity }
\end{figure}
\begin{figure}[hbt]
\includegraphics[width=1. \linewidth]{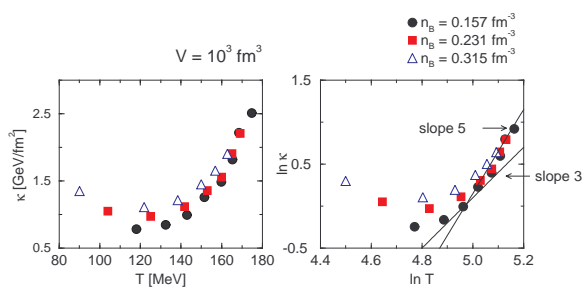}
\caption{Heat concuctivity }
\end{figure}

\begin{figure}[hbt]
\includegraphics[width=.9 \linewidth]{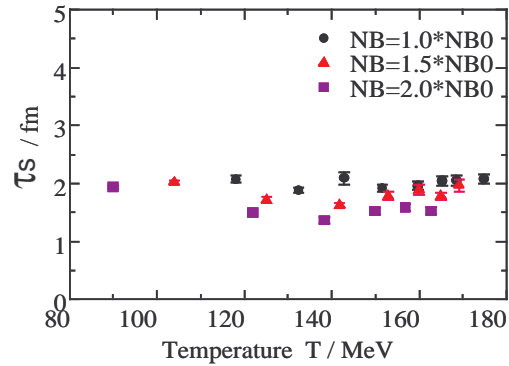}
\caption{relaxation of visous-shear tensor }
\end{figure}
\begin{figure}[hbt]
\includegraphics[width=.8 \linewidth]{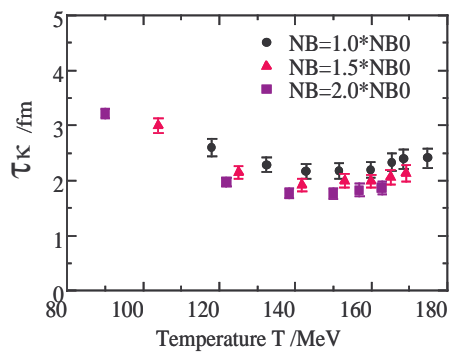}
\caption{relaxation of heat conductivety}
\end{figure}
\begin{figure}[bth]
\includegraphics[width=.9 \linewidth]{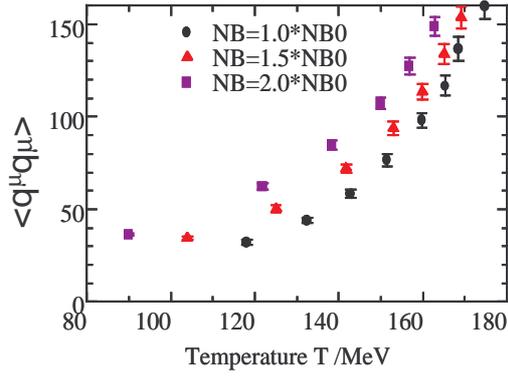}
\caption{The square expectation of heat current}
\end{figure}
\begin{figure}[hbt]
\includegraphics[width=.9 \linewidth]{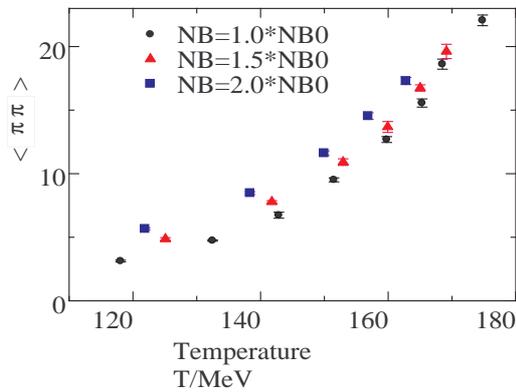}
\caption{The square expectation of visous-shear tensor}
\end{figure}


\section{Concluding remarks }

In this paper we have investigated the coefficients of the causal hydrodyanamics and evaluated them by using a hadro-molecular simulation.  According to our simulation based on URASiMA, the relaxation time of stress-shear tensor and heat current of the hadron gas are about 2 fm.  The changes with temperature and baryon number density are very small.  This results show rather clear contrast against the relaxation time in diffusion process which changes clearly with temperature $^{\cite{PRC,EPL}}$.

The importance of quantitatively investigating the relaxation in order to justify the macroscopic model has been pointed out by Iso, Mori and Namiki$^{\cite{IMN}}$.

\vspace{.5cm}

This work is supported by Grant-in-Aid for Scientific Research by 
Monbu-kagakusyo (No.\ 18540294).  Stimulating discussions at the workshop {\it ThermalField Theory  and Its Application} held at 
the Yukawa Institute for Theoretical Physics was 
extremely helpful.

\begin{small}

\end{small}

\begin{thebibliography}{99}
\bibitem{LL}L.\ D.\ Landau and E.\ M.\ Lifsitz, Fluid Mechanics(Pergamon Press, Oxford)1989.
\bibitem{NI}M.\ Namiki and C.\ Iso, Prog.\ Theor.\ Phys.\ {\bf 18} (1957), 591.
\bibitem{We}S.\ Weinberg, Astrophys.\ J.\ {\bf 168} (1971), 175.
\bibitem{IS}W.\ Israel, Ann.\ of Phys.\ {\bf 100}, 310(76);W.\ Israel and J.\ M.\ Stewart, Ann.\ of Phys.\ {\bf 118}, 341(79).
\bibitem{CH}A.\ Muronga, Phys.\ Rev.\ {\bf C69} (2004), 034903;U.\ W.\ Heinz, H.\ Song and A.\ K.\ Chaudhuri, Phys.\ Rev.\ {\bf C73} (2006), 034904;A.\ Muronga, D.\ H.\ Rischke, nucl-th/0407114;R.\ Baier, P.\ Romatschke and U.\ A.\ Wiedemann, Phys.\ Rev.\ {\bf C73} (2006),064903.
\bibitem{MS}S.\ Muroya and N.\ Sasaki, Prog.\ Theor.\ Phys.\ {\bf 113}(2005), 457.
\bibitem{Sasaki-ptp}N.\ Sasaki, Prog.\ Theor.\ Phys.\  {\bf 106} (2001), 783.
\bibitem{PRC}N.\ Sasaki, O.\ Miyamura, S.\ Muroya and C.\ Nonaka, Phys. Rev. C {\bf 62} (2000), 011901R.
\bibitem{EPL}N.\ Sasaki, O.\ Miyamura, S.\ Muroya and C.\ Nonaka, Europhys. Lett.
{\bf 54} (2001), 38. 

 \bibitem{IMN}C.\ Iso, K.\ Mori and M.\ Namiki, Prog.\ Theor.\ Phys.\
{\bf 22} (1959), 403.
\end{thebibliography}
\end{document}